# DRAMA: Commodity DRAM based Content Addressable Memory

L. Yavits

**Abstract**—Fast parallel search capabilities on large datasets provided by content addressable memories (CAM) are required across multiple application domains. However compared to RAM, CAMs feature high area overhead and power consumption, and as a result, they scale poorly. The proposed solution, DRAMA, enables CAM, ternary CAM (TCAM) and approximate (similarity) search CAM functionalities in unmodified commodity DRAM. DRAMA performs compare operation in a bit-serial fashion, where the search pattern (query) is coded in DRAM addresses. A single bit compare (XNOR) in DRAMA is identical to a regular DRAM read. AND and OR operations required for NAND CAM and NOR CAM respectively are implemented using nonstandard DRAM timing. We evaluate DRAMA on bacterial DNA classification and show that DRAMA can achieve 3.6× higher performance and 19.6× lower power consumption compared to state-of-the-art CMOS CAM based genome classification accelerator.

─────────── ◆ ───────────

## 1 INTRODUCTION

Content addressable memories (CAMs) are widely used in a variety of applications including computer microarchitecture (caches, TLBs, etc.), networking infrastructure (routers), natural language processing, databases, e-commerce, content management systems, computational biology and others. However, CAM has several major drawbacks that limit its wide adoption, especially when it comes to processing very large datasets: CAM (1) is too expensive (in terms of silicon area), (2) is too power hungry, (3) scales poorly (very large CAMs are not practical).

While recently proposed emerging memory- and 3D NAND-based solutions [3][7][10][11][13][17] may mitigate the scalability issue, they bring a host of new problems, mainly limited write endurance, high write energy consumption and high production cost. Limited write endurance means short lifetime of CAM designs [10], significantly limiting their practicality.

We propose DRAMA, a CAM implemented by commodity DRAM. We attempt to demonstrate that DRAMA resolves the aforementioned limitations. By performing search operations inside DRAM banks, in all bit-columns in parallel, DRAMA achieves very high search throughput while maintaining a typical DRAM power consumption. Through the use of DRAM infrastructure, DRAMA significantly increases CAM density and enables CAM at scale likely infeasible today.

We make the following contributions in this work:
- DRAMA enables content addressable functionality (binary CAM, ternary CAM, similarity search) in an unmodified commodity DRAM.
- We develop exact and approximate matching procedures using standard DRAM commands.

## 2 CONTENT ADDRESSABLE MEMORY

Figure 1 shows the architecture of a conventional CAM. A

• *Leonid Yavits, E-mail: leonid.yavits@biu.ac.il*
*Authors are with the Department of Engineering, Bar Ilan University, Ramat Gan, Israel.*

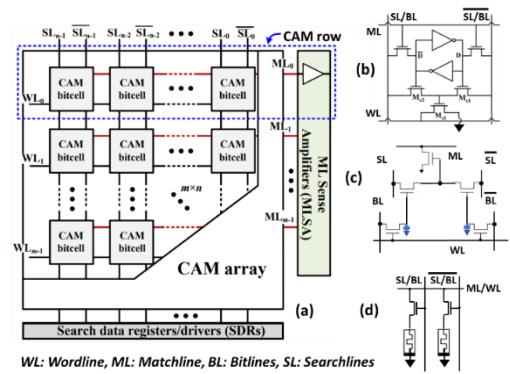

Figure 1. (a) CAM organization; NOR CAM/TCAM cells (b) CMOS static, (c) Ternary dynamic (Embedded DRAM based), (d) Ternary resistive (magnetoresistive, ferroelectric).

matchline (ML) is shared between bit cells of a row and is also fed into a matchline sense amplifier (MLSA). The bitlines (BL) and searchlines (SL) are shared across all rows of the CAM array. Read and write operations in CAM are similar to conventional RAM. Access to a bit cell is facilitated by enabling the wordline (WL) for the corresponding row and precharging or asserting the BLs/SLs for read or write operations, respectively. A search (compare) operation is performed simultaneously across the entire array during a single clock cycle, comprising two phases: (1) precharging the matchlines, (2) the evaluation which begins by asserting the query data on the SLs. The MLSAs evaluate the state of the matchlines at the end of compare cycle and signal match or mismatch.

Several CAM designs utilize emerging nonvolatile memories, such as resistive [10], magnetoresistive [3] and ferroelectric [17]. While these technologies provide higher density compared to CMOS SRAM, they suffer from high power consumption and limited write endurance. They are also slower and more expensive to manufacture than CMOS solutions.

Alternatively, 3T1C embedded DRAM based CAM designs (using CMOS [18] or tunnel field effect transistors [6]) have practically unlimited write endurance but lower density.

Recently introduced exact pattern matching accelerators

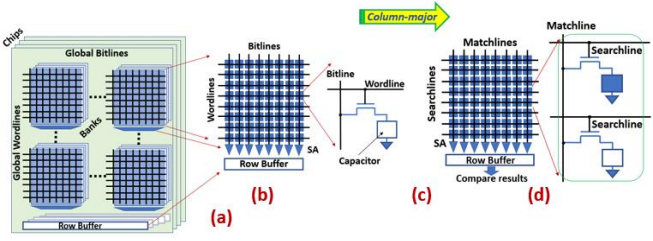

Figure 2. (a) DRAM organization, (b) DRAM subarray with row buffer, (c) DRAMA unmodified subarray, (d) DRAMA CAM cell

Sieve [16] and DRAM-CAM [15] modify DRAM by adding explicit matching logic to enable CAM functionality.

A new promising direction is 3D NAND based CAM which provides high parallelism and density [7][13]. However due to its low write endurance, 3D NAND CAM based applications are limited to those that do not require frequent database replacement or update. Additionally, supporting CAM functionality requires modifying the 3D NAND design.

## 3 DRAMA Design

DRAMA targets implementation of CAM in an unmodified commercial DRAM. Therefore DRAMA implements all operations required for exact and approximate (similarity) search using only standard DRAM commands and with no additional hardware circuitry within the DRAM.

A DRAM chip consists of multiple banks. Each bank is further divided into multiple subarrays. Figure 2(a) and (b) show DRAM organization and DRAM subarray, respectively, with the latter comprising multiple rows of DRAM cells connected to an array of sense amplifiers (SAs). Each row of DRAM cells shares a wordline that activates them. Similarly, each column of DRAM cells shares a bitline that connects those cells to the corresponding SA.

Figure 2(c) shows the DRAMA subarray. DRAM *bitlines* function as *matchlines* in DRAMA. DRAM *wordlines* become *searchlines* in DRAMA. Contrary to row-major convention, datawords in DRAMA are stored *transposed*, in column-major fashion. A CAM cell (Figure 2(d)) comprises two 1T1C DRAM cells of the same bit column. Data bit is always stored along its complement. In Figure 2(d), '1' is marked by solid color, whereas '0' is blank. DRAMA can be implemented by an unmodified off-the-shelf commercial DRAM.

### 3.1 Bitwise compare-by-read in DRAMA

DRAMA is inspired by an observation that a bitwise compare is **identical to a standard DRAM read operation**.

A compare of $m$-bit dataword to $m$-bit query can be written as follows: $Match = AND_{i=0}^{m-1}(query_i == dataword_i)$ where $query_i$ and $dataword_i$ are the $i^{th}$ bit of query and dataword respectively, "==" denotes bitwise (row-wise) XNOR.

Compare operation in DRAMA is performed in a row-by-row fashion, where only one searchline is active at a time. Figure 3 and Figure 4 show the state transitions comprising bitwise compare operation (identical to conventional DRAM read) that results in match and in mismatch, respectively.

Following the *precharge* command **PRE**, the matchline is charged at $V_{DD}/2$. The SAs are disabled and the searchlines are negated. The reference voltage of SA is also set at $V_{DD}/2$. Upon receiving the *activate* command **ACT**, the searchline corresponding to the query bit pattern of '1' is asserted. Following is the *charge sharing* phase. If the stored bit matches the query bit, the charge flows from the 1T1C cell to the matchline, raising the voltage level of the matchline to $V_{DD}/2 + \delta$ (Figure 3). If the stored bit does not match the corresponding query bit, the matchline charge flows from the matchline to the mismatching 1T1C cell, following which the matchline voltage level drops to $V_{DD}/2 - \delta$ (Figure 4). After the charge sharing is complete, the SA is enabled. It senses the difference between the matchline and the reference voltage and amplifies the deviation to drive the matchline to 0 (in the case of mismatch) or $V_{DD}$ (in the case of match). The capacitor in a cell where the searchline is asserted gets fully charged in the case of match, or fully discharged in the case of mismatch.

### 3.2 Row copy and AND/OR operations

To implement $AND_{i=0}^{m-1}$ without modifying the DRAM design, we utilize row copy and AND operations introduced in ComputeDRAM [2] and presented in Figure 5. Both operations are implemented through manipulating the DRAM timing, specifically the time intervals between PRE and ACT ($t_{RP}$) and ACT and PRE ($t_{RAS}$) outside of DRAM specifications.

For row copy **CPY(R$_T$, R$_S$)**, ACT is issued to open the source row R$_S$, followed by PRE, followed by ACT that opens the target row R$_T$. The copy is implemented by significantly shortening the $t_{RP}$ by interrupting the PRE command prematurely (Figure 5(a)).

To implement **AND/OR(R$_1$, R$_2$, R$_3$)**, the same command series is issued. However, the time intervals between commands are set to a minimum (Figure 5(b)). Since both AND and OR are based on majority operation, one of the three rows participating in AND and OR must be preset to all-zeros and all-ones respectively. There are additional constraints on the row addresses of the three argument rows R$_1$, R$_2$, and R$_3$.

### 3.3 Putting it all together

Figure 6 presents the "software-only" implementation of NAND compare operation in DRAMA. Reference D is stored in DRAM in column-major fashion, two rows per data bit (zero is coded '10', one is coded '01'). Query pattern is fed to DRAMA in a bit-serial fashion, through the **row address assignment** (data bus is not involved in transferring queries to

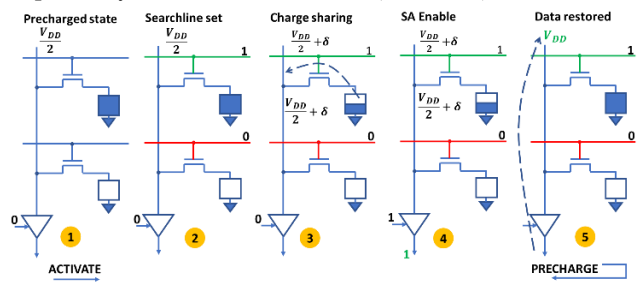

Figure 3. Match in DRAMA

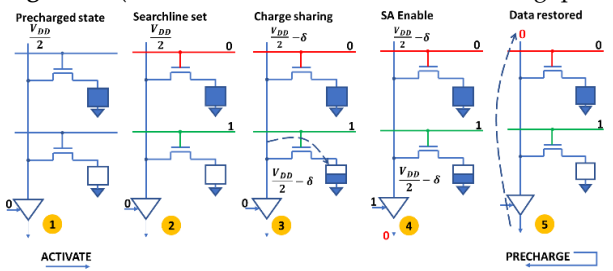

Figure 4. Mismatch in DRAMA

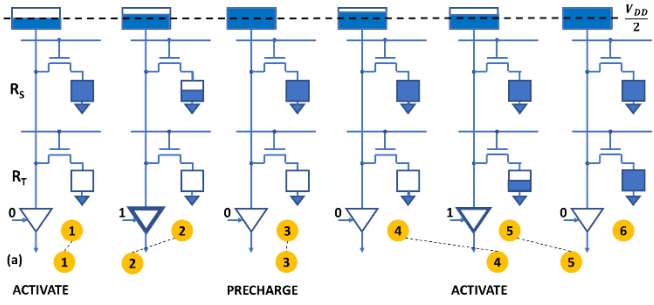

Figure 5. (a) Row Copy (b) AND, adopted from ComputeDRAM [2]

DRAMA): compare with '0' opens an even row ($Q_{2J}$); compare with '1' opens an odd row ($Q_{2J+1}$).

Figure 6(a) shows the NAND compare logic (these are not additional logic gates). DRAM rows $R_1$, $R_2$ and $R_3$ are reserved for AND operation. Rows $C_0$ and $C_1$ are prewritten with all-zeros and all-ones respectively. The "running" compare result (Match) is retained in $R_2$. Figure 6(b) shows the NAND compare pseudocode using PRE and ACT commands.

NOR CAM $Mismatch = OR_{i=0}^{m-1}(\sim query_i == dataword_i)$ can be implemented in a similar fashion with two changes: First, the query is inverted (i.e. compare with '0' opens an odd row and vice versa). Second, AND is replaced by OR. The final compare result is inverted (i.e. match is signaled by '0').

An alternative to changing DRAM timing to implement AND and OR functionality is to apply the triple row activation as suggested in Ambit [12]. However it requires a modification to DRAM, specifically to row activation logic.

The results of individual searches can be either aggregated in DRAMA using bulk logic operations or read from DRAMA (for example by an external CPU). DRAMA assumes that an external memory controller supports user-configurable DRAM timing which might require its modification.

### 3.4 Support for Ternary CAM (TCAM) functionality

DRAMA supports TCAM functionality by coding the "don't care" bit as '11' for NAND ('00" for NOR TCAM). In such case, the matchline level is pulled up towards $V_{DD}/2 + \delta$ (pulled down towards $V_{DD}/2 - \delta$) during the charge sharing regardless of the query bit value ('01' or '10'). Consequently, such per-row compare makes no impact on the final compare result $Match \cap 1 = Match$ ($Mismatch \cup 0 = Mismatch$).

### 3.5 Impact on refresh

During a compare cycle, a matching bit is refreshed. Since a bit compare has five-six orders of magnitude lower latency (<200 ns) compared to the refresh interval (64 ms), it is reasonable to assume that during a refresh interval, every bit will match and mismatch at least once. To increase the probability even further, NOR and NAND CAM operations can be alternated. This will ensure that if a specific bit becomes "sticky" and never matches in NAND compare, it will necessarily match in NOR compare. When compare operations are not conducted, a conventional refresh can be applied.

### 3.6 Approximate Search

While CAMs are typically designed to find exact matches, approximate or similarity search is increasingly required by many contemporary applications such as genome analysis. In approximate search, if the difference (for example Hamming distance) between a stored reference dataword and the query is below a certain predefined threshold, such stored dataword is considered a "match".

Figure 7(a) shows the logic (not a hardware implementation) of a NAND compare based approximate search. The Hamming distance (HD) tolerated in this search is 1. Figure 7(b) presents the pseudocode of the approximate search.

## 4 EVALUATION

We use bacterial genome classification for functional evaluation of DRAMA. The bacterial database [8] is stored in DRAMA, multiple *kmers* (DNA fragments of *k* DNA bases) per column, in column-major format, using one-hot encoding. Specifically, each DNA base is stored in 4 vertically adjacent DRAM cells as follows: A=0001, G=0010, C=0100, T=1000. While one-hot occupies the same number of cells as the basic two CAM cells per base format, it allows a single DRAM read per DNA base compare. This is because only one out of four rows storing a DNA base is activated in each such compare. The reference kmers of each distinct species are assigned to a certain (known in advance) DRAM column group, to enable an efficient identification of the species through column address.

Genome classification is composed of two steps: (1) query kmer search, (2) the taxon assignment, implemented by external CPU. Specifically, after the search result is fetched to CPU, the taxon is inferred from the column address of a data item where "1" (match) is located.

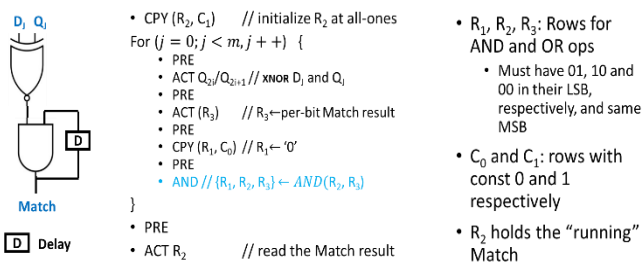

Figure 6. NAND Compare (a) Compare logic (this is not a hardware implementation) (b) Pseudocode that implements the logic.

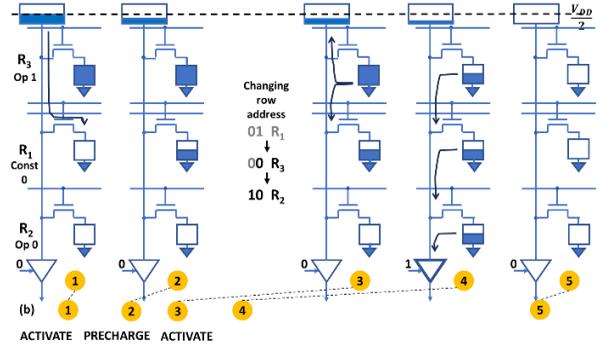

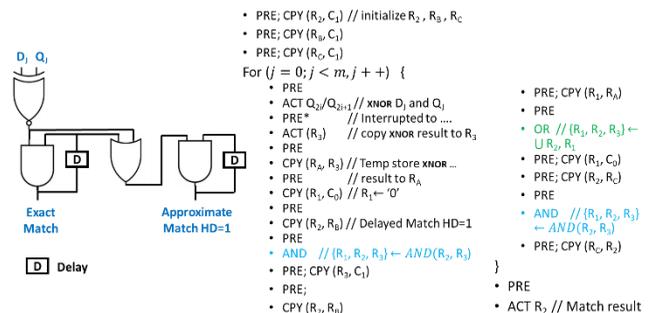

Figure 7. Approximate (Hamming distance of 1 tolerant) compare (a) Compare logic (this is not a hardware implementation) (b) Pseudocode.

We verify DRAMA's functionality and performance using a custom trace-driven simulator. We use Ramulator [9] and Vampire [5] for power estimation. We model DRAMA as 16-chips 8-bank/chip 128×64 columns/bank DD3 (no subarray level parallelism) and compare it with the following state-of-the-art solutions:

- Baseline CPU (Core i7-12700K, L1=256KiB, L2=2MiB, L3=25MiB, 64GB DRAM) running classification tool Kraken2 [14].
- Sieve [16], a DRAM based platform for kmer matching. Sieve stores the reference genome database, the queries and the taxa data in DRAM and compares kmers in bit-serial fashion using specially designed matching hardware placed within DRAM banks. We assume Type-2 Sieve mode (no subarray level parallelism), the same chip and bank configuration as DRAMA.
- DASH-CAM [18], an embedded dynamic CAM based pathogen classification accelerator (64K×32-mer at 1GHz) which performs bit-parallel compare operations. Since the entire reference bacterial dataset does not fit in DASH-CAM, it needs to be uploaded and searched in chunks. Following Sieve, the query batch size is set at 64.

*Throughput* in G*kmers*/sec is presented in Table 1. The throughput of the baseline CPU diminishes with the dataset size, with dips at points when the dataset no longer fits in a particular level of cache. DASH-CAM underperforms both Sieve and DRAMA. However, if the query batch is significantly increased, DASH-CAM performance may improve considerably due to its parallel operation. Sieve slightly outperforms DRAMA because it implements matching in hardware.

*Power* consumption figures are presented in Table 1. For the baseline CPU, we do not account for the CPU "housekeeping". However, we do account for DRAM access power for the baseline CPU, DASH-CAM and DRAMA.

The delay and energy consumption division between kmer search and taxa assignment in DRAMA is 94%/6% and 85%/15% respectively.

Table 1. Throughput and power consumption

| Platform | CPU | DASH-CAM | Sieve | DRAMA |
| --- | --- | --- | --- | --- |
| Throughput, Gkmers/s | 0.4-0.25 | 45 | 164 | 149 |
| Power, W | 0.3 | 5.7 | 0.27 | 0.29 |

*Area*. Similar to ComputeDRAM, DRAMA requires no DRAM modifications but stores every bit using two DRAM cells. The matching logic and temporary storage in Sieve claims between 1% and 10.75% of area overhead. Sieve additionally stores queries alongside reference data (which may occupy 12.5% of the subarray), as well as the taxa data.

DASH-CAM is based on an embedded dynamic 3T1C cell, featuring relatively low density, which requires searching the database in chunks, ultimately limiting the performance.

DRAMA enables approximate search (i.e. search that tolerates Hamming distance of 1 between query and reference datawords) which typically improves the classification accuracy [16]. Since both Sieve and Kraken2 do not support approximate search, we do not include it in our evaluation.

## 5 CONCLUSIONS

We present DRAMA, a commodity DRAM based Content Addressable Memory (CAM), supporting CAM, ternary CAM and approximate (similarity) search capabilities. Both NAND and NOR CAM implementations are developed and presented. Compared with CMOS as well as emerging memory technologies (resistive, magnetoresistive, and ferroelectric), DRAMA provides highly dense CAM solution, that can be efficiently scaled to support very large (potentially hundreds of Gigabytes) searchable datasets. In contrast with DRAM based outperforming state-of-the-art solutions, DRAMA can be designed around an unmodified off-the-shelf commodity DRAM. Compared with 3D NAND based CAM, DRAMA is not restricted by write endurance and hence is not limited to applications which do not require frequent database replacement or updates. DRAMA allows very high parallelism, which translates into very high throughput and energy efficiency.

## ACKNOWLEDGMENT

This work is partially supported by the European Union's Horizon programme for research and innovation [101047160 - BioPIM], and in part by the Israeli Ministry of Science and Technology under Lise Meitner grant No. 1001569396 for Israeli-Swedish research collaboration.